\documentclass[epjST]{svjour}
\usepackage{graphics,dsfont,color}
\def\one{{\mathchoice {\rm 1\mskip-4mu l} {\rm 1\mskip-4mu l} {\rm
1\mskip-4.5mu l} {\rm 1\mskip-5mu l}}}

\def\tr{{\rm tr}\; }

\newcommand{\ket}[1]{| #1 \rangle}
\newcommand{\bra}[1]{\langle #1 |}

\newcommand{\inner}[2]{ \langle #1 | #2 \rangle}

\newcommand{\bitem}{\begin{itemize}}
\newcommand{\eitem}{\end{itemize}}
\newcommand{\benum}{\begin{enumerate}}
\newcommand{\eenum}{\end{enumerate}}
\newcommand{\beq}{\begin{equation}}
\newcommand{\eeq}{\end{equation}}
\newcommand{\beqa}{\begin{eqnarray}}
\newcommand{\eeqa}{\end{eqnarray}}
\newcommand{\bproof}{\begin{proof}}
\newcommand{\eproof}{\end{proof}}
\newcommand{\bprop}{\begin{proposition}}

\newcommand{\bdef}{\begin{definition}}

\def\openone{\leavevmode\hbox{\small1\kern-3.8pt\normalsize1}}%

\begin{document}
\title{Fast Quantum Methods for Optimization}
\author{ Sergio Boixo\inst{1} \and Gerardo Ortiz\inst{2} \and Rolando Somma\inst{3} 
\thanks{\emph{Rolando Somma, Los Alamos National Laboratory,
MS B213, Los Alamos, NM 87545, USA}}%
}                     
\offprints{}          
\institute{Google Quantum A.I. Labs, Venice, CA 90291, USA \and Department of Physics, 
Indiana University, Bloomington,
IN 47405, USA \and Theoretical Division, Los Alamos National Laboratory,
Los Alamos, NM 87545, USA}
\date{Received: date / Revised version: date}
%
\abstract{ Discrete combinatorial optimization consists in finding the
  optimal configuration that minimizes a given discrete objective
  function.  An interpretation of such a function as the energy of a
  classical system allows us to reduce the optimization problem into
  the preparation of a low-temperature thermal state of the
  system. Motivated by the quantum annealing method, we present three
  strategies to prepare the low-temperature state that exploit quantum
  mechanics in remarkable ways. We focus on implementations without
  uncontrolled errors induced by the environment. This allows us to
  rigorously prove a quantum advantage.  The first strategy uses a
  classical-to-quantum mapping, where the equilibrium properties of a
  classical system in $d$ spatial dimensions can be determined from
  the ground state properties of a quantum system also in $d$ spatial
  dimensions. We show how such a ground state can be prepared by means
  of quantum annealing, including quantum adiabatic evolutions.  This
  mapping also allows us to unveil some fundamental relations between
  simulated and quantum annealing.  The second strategy builds upon
  the first one and introduces a technique called spectral gap
  amplification to reduce the time required to prepare the same
  quantum state adiabatically. If implemented on a quantum device that exploits
  quantum coherence, this strategy leads to a quadratic improvement in
  complexity over the well-known bound of the classical simulated
  annealing method. The third strategy is not purely adiabatic;
  instead, it exploits diabatic processes between the low-energy
  states of the corresponding quantum system.  For some problems it
  results in an exponential speedup (in the oracle model) over the
  best classical algorithms. }
%
%
\maketitle
%

\section{Introduction}
\label{somma:sec:introduction}
Discrete combinatorial optimization problems are ubiquitous in science and technology but 
often hard to solve~\cite{somma:optimizationbook}. The main goal is to find the (optimal) configuration
that corresponds to a global minimum of a given objective function. As the dimension of the search
space   typically grows exponentially with the size of the problem, finding the optimal configuration
by exhaustive search rapidly  becomes  computationally intractable.
This is the case even for relatively small
problem sizes, specified by fifty or more bits. Efficient  strategies for optimization
are highly desirable.

Historically, some of the most practical optimization methods were developed
in the context of physics simulation. Simulated annealing (SA)~\cite{somma:simulatedannealing}, for example,
is a well-known method that imitates the cooling process of a classical system (e.g., a metal) that is initially
heated and then cooled slowly,
so that it ends up in one of the lowest-energy configurations. Ideally,
 the final state of the system is represented by the low-temperature Gibbs (equilibrium) distribution.
  To solve a combinatorial
optimization problem with SA, the objective function is interpreted as the energy of the 
system. The annealing process can be simulated
on a conventional (classical) computer by means of probabilistic Monte Carlo 
methods~\cite{somma:montecarlobook}. 
Such a process is determined by a sequence of transition rules, or stochastic matrices, that depend
 on a parameter associated with the (inverse) temperature of the system. In some cases, a good choice
 of transition rules allows us to sample an optimal configuration
 and solve the combinatorial optimization problem, with
 high probability, using significantly less resources than exhaustive search.
 SA can be applied to a variety of difficult problems,
 such as the well-known traveling salesman problem (TSP)~\cite{somma:TSP}
 or Ising spin glasses~\cite{somma:SAIsingGlass}.
 
 Quantum mechanics provides remarkable tools for problem solving~\cite{somma:shoralgorithm},
 and further motivates the search of novel and fast algorithms for optimization.
 A natural approach to develop such algorithms follows by considering a ``quantum
 version'' of SA. Rather than preparing a low-tempera\-ture Gibbs distribution of a classical system,
 the goal is now to prepare the lowest-energy or ground state of a {\em quantum} system.
 A projective measurement of the ground state would allow us to sample
 the optimal configurations, and solve the problem, with high probability.
 Such is the basic idea behind quantum annealing (QA)~\cite{somma:QA1,somma:QA2,somma:QA3}, adiabatic
 quantum computation (AQC)~\cite{somma:AQC1}, or general quantum adiabatic state transformations~\cite{somma:AST1}.
 These methods specify a   quantum evolution to prepare the desired quantum state.
 The evolution can be simulated on a conventional computer by classical algorithms (e.g., by using
 quantum Monte Carlo methods or by numerical simulation of Schr\"odinger's differential equation)~\cite{somma:QA4},
 on a quantum computer by quantum algorithms~\cite{somma:AST1}, implemented directly with quantum
 simulators (c.f.,~\cite{somma:simulator}), or with programable
 physical QA architectures~\cite{somma:qa108,somma:shin,somma:definingSpeedup}. 
 The complexity of each simulation or implementation may be different.
 The power of QA has been studied in a number of examples, such as in finding the low-energy
 configurations of Ising spin glasses~\cite{somma:spinglass,somma:qamagnet}.
 
 This paper considers QA based methods for optimization and studies
 the complexity of such methods in the context of quantum computation
 or quantum simulation. Our QA strategies require a device that uses
 quantum coherence, such as a quantum computer or quantum simulator, for their
 implementation.  Contrary to other results on the power of QA, which
 are commonly suggested from numerical evidence, our aim here is to
 mathematically prove that QA can outperform classical algorithms for
 some optimization problems.  We will then focus in closed-system QA
 without uncontrolled errors induced by the environment, where the
 state of a quantum system generally evolves according to the
 Schr\"odinger equation, with the only condition that the final state
 is (close to) the appropriate eigenstate  of the corresponding quantum system.
 In particular, AQC is one type of QA, with the additional constrain
 that the quantum evolution is adiabatic
 \cite{somma:bohm51,somma:RigoOrt}, so that the quantum state at any
 time is (close to) the ground state of the perturbed
 system\footnote{Other
   continuous-time evolutions  exploit diabatic
   transitions to prepare the ground state, such as those based on ``shortcuts to
   adiabaticity''~\cite{somma:shortcuts}, and can  be considered as a type of QA.
  }.  A main
 goal of this paper is to theoretically and rigorously prove that {\it
   quantum implementations} of closed-system QA can be significantly
 more powerful than SA for solving optimization problems.  In
 contrast,
 Refs.~\cite{somma:qa108,somma:definingSpeedup,somma:qamagnet} study
 QA in the context of open-system dynamics.

 We study three different closed-system QA strategies.  The first
 strategy considers a classical-to-quantum mapping, where a stochastic
 matrix is transformed into a Hamiltonian $H_\beta$ that models a
 quantum system \cite{somma:qmap1,somma:qmap2,somma:qmap2p}. The
 ground state of this Hamiltonian is related to the stationary state
 of the stochastic matrix. That is, measurements of the ground state
 in the computational basis produces configurations according to the
 Gibbs distribution of the corresponding classical system.  We then
 study the complexity of different QA techniques to prepare the ground
 state of $H_\beta$ and compare it with the complexity of SA; QA and
 SA could have similar complexities in this case.  The second strategy
 improves upon the first one and constructs a different Hamiltonian
 $\tilde H_\beta$ based on the idea of ``spectral gap
 amplification''~\cite{somma:qsofsa,somma:qmap3,somma:gapamplification}.
 $\tilde H_\beta$ also has the ground state of $H_\beta$ as eigenstate
 (not necessarily the ground state), so that QA can be used to prepare
 the eigenstate.  We will prove that the complexity of preparing such
 a state, on a quantum device, is of order $1/\sqrt{\Delta}$, where
 $\Delta$ is a lower bound on the spectral gap of the stochastic
 matrix. This represents a quadratic quantum speedup with respect to
 SA, where the complexity is of order
 $1/\Delta$~\cite{somma:stroockbound}. 
 (Typically, $\Delta \ll 1$ in hard instances of optimization
 problems.)  The third strategy exploits diabatic transitions to
 excited states to prepare the ground state of the quantum
 system~\cite{somma:expspeedup}. We show that this approach is
 efficient in solving a particular oracular
 problem~\cite{somma:gluedtrees}, even though the minimum energy gaps
 are exponentially small in the problem size. In contrast, classical
 algorithms require (provable) exponential time in this case.  We also
 discuss generalizations of this QA approach based on initial-state
 randomization to avoid some slowdowns due to small gaps, and comment
 on recent results on MAX 2-SAT, a NP-hard satisfiability
 (optimization) problem investigated in Ref.~\cite{somma:max2sat}.
  
  Each strategy is explained in detail in Secs.~\ref{somma:sec:2},~\ref{somma:sec:3}, and~\ref{somma:sec:4}, respectively. 
  We summarize the results and conclude in Sec.~\ref{somma:sec:conclusions}.


\section{Quantum vs. simulated annealing}
\label{somma:sec:2}
Is QA fundamentally different from SA? Is there any reason why QA should outperform SA
in solving optimization problems? In this section, we try to address these questions
and explore some fundamental connections between SA and QA following
Refs.~\cite{somma:qmap3,somma:qmap1,somma:qmap2,somma:qmap2p}. 

We first provide a short summary of the SA method for optimization.
The search space of the combinatorial optimization problem, $\Sigma= \{ \sigma_0, 
\ldots, \sigma_{N-1} \}$, consists of $N$ configurations $\sigma_i$; the goal of 
SA is to find the (optimal) configuration that minimizes a given
objective function $E : \Sigma \rightarrow \mathds R$. Typically, each configuration $\sigma_i$
can be associated with a state of a classical system defined
on a lattice (or graph) $\Lambda$ of size $n$, so that $E(\sigma_i)$ is the energy of the state.
Each vertex of the lattice is a ``site'' of the classical system, and each site can be in one
of $M$ possible states. For example, if configurations are represented by $n$-bit strings 
(i.e., $N=2^n$), the classical system may correspond to a classical Ising model of $n$ sites, and $M=2$.
The spatial dimension of $\Lambda$ is $d$.

Monte Carlo implementations of SA 
generate a stochastic sequence of configurations, via a sequence of Markov processes, 
that converges to the low-temperature Gibbs (probability) distribution
defined by the probabilities
\begin{equation}
\pi_{\beta_m} (\sigma_i)\propto \exp (-\beta_m E(\sigma_i)) \; . 
\end{equation}
When $\beta_m$ is sufficiently large, sampling from the Gibbs distribution
outputs the lowest-energy state of the system (i.e., an optimal configuration) with large probability. 
The  annealing  process is determined by a sequence of 
$N \times N$ stochastic matrices (transition rules)
$S_{\beta_1}$, $ S_{\beta_2}, \ldots, S_{\beta_m}$. The  real  parameters $\beta_j$
denote  a  sequence  of  ``inverse temperatures".
If $|\beta_{k+1} -\beta_k|$ is sufficiently small, the annealing process
drives the Gibbs distribution from the initial $\beta_0=0$ (i.e., the uniform distribution)
towards the Gibbs distribution for $\beta=\beta_m$~\cite{somma:stroockbound}. 
The stochastic matrices are chosen so that ${\pi}_\beta = 
(\pi_\beta(\sigma_0),\ldots,\pi_\beta(\sigma_{N-1}))^{\sf T}$ is the unique
fixed point of $S_\beta$ (i.e., $S_\beta \pi_\beta = \pi_\beta$), for all $\beta$. ($\sf T$ is the transpose.)

\subsection{Classical-to-quantum mapping}
\label{somma:sec2.1}
Remarkably, any classical 
system on a lattice $\Lambda$ can be  
related to a quantum system defined on the same lattice. 
To realize this ``mapping'', we associate
each configuration or state $\sigma_i$  with a quantum state
$\ket{\sigma_i}$. Then, $\{\ket{\sigma_0}, \ldots , \ket{\sigma_{N-1}} \}$ forms a basis of 
a Hilbert space (i.e., the computational basis). In  
this basis,  the objective function or energy functional $E$ maps to a diagonal
Hamiltonian matrix $\hat E$, where each diagonal entry is the corresponding  
$E(\sigma_i)$.  For example, for classical Ising (spin-1/2) models, each configuration 
$\sigma_i$ can be represented by  the string 
  $\sigma_i \equiv \sigma^0_i \ldots \sigma^{n-1}_i$, where $\sigma^l_i = \pm 1$.
Then, $\hat E$ can be written in operator form
by replacing $\sigma^l _i \rightarrow {\sigma}_{\bf z}^l$ in $E$, where
$\sigma_{\bf z}^l$ is the Pauli operator ($N \times N$ matrix) acting on the $l$-th site. 
For instance, quadratic objective functions are mapped to Ising models:
\begin{equation}
\label{somma:eq:quadraticE}
\hat{E}=\sum_{l=0}^{n-1} h_l\sigma_{\bf z}^l  + \sum_{l,l'=0}^{n-1} J_{ll'} \sigma_{\bf z}^l \sigma_{\bf z}^{l'} \; ,
\end{equation}
where $h_l$ denotes the local field of the spin at site $l$ and $J_{ll'}$
denotes the Ising interactions between spins at sites $l$ and $l'$.

In thermal equilibrium, the expectation
value of a thermodynamic variable $A$ in the canonical (Gibbs) ensemble is given
by
\begin{equation}
\label{somma:eq:correl1}
\langle A \rangle_\beta ={\cal Z}_{\beta}^{-1}
\sum \limits_{  \sigma_i \in \Sigma  } e^{ - \beta E({ \sigma_i  })}
A( \sigma_i  ),
\end{equation}
where ${\cal Z}_{\beta} = \sum _{  \sigma_i  \in \Sigma } e^{ - \beta E(
\sigma _i)}$ is the partition function. Note that $\pi_\beta(\sigma_i) = {\cal Z}_{\beta}^{-1} e^{ - \beta E(\sigma_i )}$
is the Gibbs distribution and $\langle A \rangle_\beta = \sum_{\sigma_i \in \Sigma}\pi_\beta(\sigma_i)A( \sigma_i  )$
is the corresponding average.
The mapping between classical and quantum states
also allows us to define $\hat A$, a diagonal $N \times N$ matrix or operator that 
has $A(\sigma_i)$ as diagonal entries. Then, 
we can rewrite  Eq. (\ref{somma:eq:correl1}) as
\begin{equation}
\label{somma:eq:correl3}
\langle A
\rangle_\beta \equiv \langle \hat{A} \rangle_\beta =  
 \bra{\psi_{\beta}} \hat{A} \ket{\psi_{\beta}} \; ,
\end{equation}
where $\ket{\psi_\beta}$ is the (normalized)
quantum state $\ket{\psi_{\beta}} = \sum_{\sigma_i \in \Sigma} \sqrt{\pi_\beta(\sigma_i)} \, \ket {\sigma_i}$.
A projective quantum measurement of $\ket{ \sigma_i}$ in $\ket{\psi_{\beta}}$ outputs the configuration 
$\sigma_i$ with probability according to the Gibbs distribution.

The state $\ket{\psi_\beta}$ can be shown to be the unique ground state
of certain quantum systems whose Hamiltonians are
defined on the same lattice
$\Lambda$~\cite{somma:qmap1,somma:qmap2,somma:qmap2p,somma:qmap3,somma:qmap4,somma:qmap5,somma:qmap6}. Assume that
the stochastic matrix $S_\beta$
with unique fixed point $\pi_\beta$  
satisfies the ``detailed balance condition'' (DBC)
\begin{equation}
    \Pr \!_\beta(\sigma_i|\sigma_j) \pi_\beta(\sigma_j) = \Pr \!_\beta(\sigma_j|\sigma_i) \pi_\beta(\sigma_i)\;,
\end{equation}
where $\Pr_\beta(\sigma_i|\sigma_j)$ is the conditional probability
that specifies the $(i,j)$ entry of $S_\beta$. Then, we define $H_\beta$ via
\begin{equation}
\label{somma:eq:QCM}
\bra{\sigma_i} H_\beta \ket{\sigma_j} = \delta_{i,j} - \sqrt{\Pr \!_\beta(\sigma_i|\sigma_j) \Pr\!_\beta(\sigma_j|\sigma_i)} \; .
\end{equation}
Using the DBC, a simple analysis shows that $H_\beta \ge 0$ and 
$H_\beta \ket{\psi_\beta}=0$~\cite{somma:qmap1,somma:qmap2,somma:qmap2p,somma:qmap3}.
Furthermore, $H_\beta$ is irreducible when $S_\beta$ is,  and $\ket{\psi_\beta}$
is the unique ground state of $H_\beta$ in this case. The eigenvalues of $H_\beta$ are $1-\lambda_i$,
where $\lambda_i$ are the eigenvalues of $S_\beta$.

We now provide a specific construction for a Hamiltonian $H_\beta$, following Refs. \cite{somma:qmap1,somma:qmap2,somma:qmap2p,somma:qmap3},  
that provides  insight into the connections between SA and QA
-- see Sec.~\ref{somma:sec2.2}.
For illustrative purposes,
we consider again the classical Ising model and map it to a quantum
Ising (spin-1/2) model. For simplicity, we assume that $S_\beta$ is obtained
via a very similar process than that for the so-called Metropolis-Hastings algorithm
(or Glauber dynamics). 
If $\sigma_i$ and $\sigma_j$
differ in a single position (spin flip), we choose 
\begin{equation}
\Pr \!_\beta (\sigma_i|\sigma_j) = \chi \exp \{\beta[E(\sigma_i)-E(\sigma_j)]\} \; .
\end{equation}
The constant of proportionality can be set to 
\begin{equation}
\chi = \exp (-\beta \kappa )/n \; ,
\end{equation}
where $\kappa = \max_{i,j} |E(\sigma_i)-E(\sigma_j)|$, and the maximum is taken over those $i,j$
such that $\sigma_i$ and $\sigma_j$ differ by a single spin flip.
This choice guarantees that the conditional probabilities are properly bounded.
 If $\sigma_i$ and $\sigma_j$ differ in two or more positions, 
\begin{equation}
\Pr\! _\beta (\sigma_i|\sigma_j)=0 \; .
\end{equation}
Further, normalization implies 
\begin{eqnarray}
\Pr \!_\beta(\sigma_j| \sigma_j) = 1- \sum_{\sigma_i \ne \sigma_j} \Pr \!_\beta(\sigma_i|\sigma_j) \; .
\end{eqnarray}
The classical-to-quantum mapping of Eq.~(\ref{somma:eq:QCM}) gives
 \begin{equation}
 \label{somma:eq:Qmetropolis}
 H_\beta = \sum_{l=0}^{n-1} \chi \exp(\beta \hat E^l)- \chi \, \sigma_{\bf x}^l \; ,
 \end{equation}
 where $\hat E^l$ is the diagonal matrix obtained as
\begin{eqnarray}
\hat E^l = (\hat E - \sigma_{\bf x}^l \, \hat E \, \sigma_{\bf x}^l)/2 \; ,
\end{eqnarray}
and $\sigma_{\bf x}^l$ is the Pauli ``spin-flip operator'' at site $l$. 
That is,
\begin{eqnarray}
\nonumber
\sigma_{\bf x}^l = \one_2 \otimes \cdots \otimes \one_2   \otimes \! \! \! \! \! \!
\underbrace{{\bf \sigma_x}}_{l-\!{th} \ {\rm position}} \! \! \! \!  \! \! \otimes \one_2 \cdots \otimes
\one_2  \; , \ {\bf \sigma_x}= \pmatrix{ 0&1 \cr 1&0 \cr}  , 
\end{eqnarray}
and $\one_2$ is the $2 \times 2$ identity matrix.
In other words, $\hat E^l$ includes only those terms in $\hat E$ that
contain $\sigma_{\bf z}^l$; e.g., for Eq.~(\ref{somma:eq:quadraticE}),  
\begin{equation}
\hat E^l = \sigma^l_{\bf z} (h_l + \sum_{l' \ne l} J_{ll'}  \sigma^{l'}_{\bf z} )\; .
\end{equation}
Simple inspection shows that 
$(\exp(\beta \hat E^l)- \sigma_{\bf x}^l) \ket{\psi_\beta}=0$ for all $l$, 
and $(\exp(\beta \hat E^l)-\sigma_{\bf x}^l) \ge 0$~\cite{somma:qmap1}.
These properties imply that $H_\beta$ is a so-called ``frustration-free Hamiltonian'' -- see Sec.~\ref{somma:sec:3}.
We remark that the range of interactions in $H_\beta$ is determined by that of $\hat E^l$ or, equivalently, $\hat E$.

We emphasize the simplicity of Eq.~(\ref{somma:eq:Qmetropolis}): The thermodynamic
properties of any finite two-level (spin-1/2) classical system at nonzero temperature can be
obtained by computing the ground state properties of  a spin-1/2
quantum system, whose   interactions depend on
 $\beta$ and $E$, and an external and
homogeneous transverse field.  Remarkably, this field generates quantum
fluctuations that are in one-to-one correspondence with the classical
fluctuations at the inverse temperature $\beta$.  In particular, 
$H_{\beta=0} = \one_{2^n}- \sum_{l=0}^{n-1} \sigma_{\bf x}^l/n$,  thus its ground state has
all spins aligned along the external field, i.e. $\ket{\psi_{\beta=0}} =
\sum_{\sigma_i \in \Sigma} \ket{\sigma_i}/\sqrt {2^n}$. This quantum state  can be identified with
the completely mixed state (uniform distribution) of the classical
model at infinite temperature. 
In general, the low-temperature limit is 
$\ket{\psi_{\beta \rightarrow \infty}} \rightarrow \sum_{\sigma_i \in \Sigma_0} \ket {\sigma_i}/\sqrt{|\Sigma_0|}$,
where $\Sigma_0 \subseteq \Sigma$ is the set of optimal configurations or states that minimize $E$.

As shown in 
Ref. \cite{somma:qmap1}, the above classical-to-quantum  
 mapping can be realized in any finite-dimensional 
classical system in addition to Ising-like (two states) or spin-1/2 models. In Ref. 
\cite{somma:qmap1}, we illustrated this generality by mapping a classical 
(three-state) Potts model into its corresponding quantum version on the 
same lattice.

\subsection{State preparation and rates of convergence}
\label{somma:sec2.2}
The fact that $\ket{\psi_\beta}$ is the unique ground state
of a simple Hamiltonian motivates the development of quantum
methods to prepare it. The most natural methods in this context are those 
based on QA, including the recently developed methods
for quantum adiabatic state transformations in Refs.~\cite{somma:AST1,somma:AST2},
whose complexities can be shown to be smaller than that determined by standard 
adiabatic approximations~(c.f., \cite{somma:adiabaticapprox}).
Such methods require a quantum computer or simulator for their implementation.

Similar to SA, one QA method considers changing
$\beta$ slowly from $\beta_0=0$ to $\beta_m$
in the Hamiltonian $H_\beta$ of Eq.~(\ref{somma:eq:QCM}). 
In contrast to SA, this evolution is coherent and $\beta$
does not correspond to the actual inverse temperature of the quantum
system, which is assumed to be at zero temperature at all times (i.e., $\beta$ is a 
parameter that determines the strengths of the interactions in the quantum system).
 If the system is closed and the evolution is  adiabatic \cite{somma:bohm51,somma:RigoOrt},
the state of the system remains sufficiently {\em close}  to the  ground
state $\ket{\psi_{\beta}}$ along the path, transforming $\ket{\psi_{\beta_0}}$
into the desired state $\ket{\psi_{\beta_m}}$. (Usually,
$\ket{\psi_{\beta_0}}$ can be prepared easily.)

The total evolution time of the above QA method is determined by the
quantum adiabatic approximation and  depends on the spectral
gap $\Delta_{\beta}$ of $H_{\beta}$, which is the difference between
the two lowest eigenvalues of $H_{\beta}$ (i.e., the first positive
eigenvalue in this case).  For arbitrary precision, such a gap
determines a bound on the rate at which $\beta$ can be
increased. Assuming $E = O(n)$, which is typical in, e.g., Ising
models, the gap satisfies $\Delta_{\beta} = \Omega (\exp(-\beta p
n))$, where $p>0$ is a constant~\cite{somma:qmap1}.  This lower bound
was determined using the inequalities in Ref.~\cite{somma:hopf63}. It
is based on the worst-case scenario, so it is expected to be improved
on a case-by-case basis by exploiting the structure of $H_\beta$ (or
$S_\beta$). According to the folk version of the quantum adiabatic
approximation~\cite{somma:bohm51,somma:qmap1,somma:qmap2,somma:qmap2p,somma:qmap3}, the rate $\dot
\beta(t)$ can be determined from
\begin{equation}
\label{somma:eq:adiabcon}
\frac{ \| \partial_{\beta} \ket{ \psi_{\beta}} \| \dot \beta(t)
}{\Delta_{\beta(t)}} \le \epsilon ,  \ 0  \le t \le {\cal T},
\end{equation}
where $\epsilon$ is an upper bound to the error probability and $\cal T$ is the total time
of the evolution.  Note that  $ \| \partial_{\beta} \ket{
\psi_{\beta}} \|\le   \| (\langle \hat{E}
 \rangle_{\beta} - \hat E) \ket{ \psi_{\beta}} /2 \| \le E_{\max}$,
where $E_{\max} \ge \max_{\sigma_i} |E(\sigma_i)|$ is an upper bound on the objective function. 
Thus, $ \| \partial_{\beta} \ket{ \psi_{\beta}}\|$
$= O(n)$ under the assumption on $E$.

In the limit  $\log t \gg n \gg 1$, and for constant and small $\epsilon$,
integration of Eq.~(\ref{somma:eq:adiabcon}) with the bound $\Delta_{\beta} = \Omega (\exp(-\beta p n))$ gives
\begin{equation}\label{somma:eq:geman_geman}
\beta(t) = O(\log t/n)\;.
\end{equation}
This result is somewhat similar to the Geman-Geman asymptotic convergence rate  for SA obtained in
Ref.~\cite{somma:gem84}. Such an agreement results
from the fact that $\Delta_\beta$ is also the gap of $S_{\beta}$ (see Sec.~\ref{somma:sec2.1}), and the
   complexity  of SA can be shown to be  of order
   $1/\Delta_\beta$~\cite{somma:stroockbound}. 

The total evolution time can be determined from $\beta({\cal
  T})=\beta_m$. To obtain the optimal
configuration with high probability from the Gibbs
distribution, it suffices to choose
$\beta_m = O(n)$ in the worst case. This gives a total
time ${\cal T}$ or complexity for the current QA method  (and similarly for SA) of order $\exp(c_1 n^2)$, for some constant
$c_1>0$. This bound in complexity is
larger than that of exhaustive search (which is exponential in $n$), but it is an absolute worst case bound and SA is
performed much faster in practice. For example, in many cases it suffices
to choose $\beta_m = O(\log n)$, leading to a ${\cal T} = \exp(c_2 n \log n)$, for some constant
$c_2>0$.


As Eq.~(\ref{somma:eq:adiabcon}) does not necessarily constitute a necessary {\it and} 
sufficient condition for the validity of the adiabatic theorem \cite{somma:RigoOrt}, 
 we explain next a version of the adiabatic approximation
obtained by evolution
randomization~\cite{somma:AST2,somma:phaserandom2}. We discretize the
path into $m$ steps with separation $(\beta_{r+1} -\beta_r ) = \Omega(
1/n)$, and at the $r$th step we perform an evolution $e^{i 
  H_{\beta_r} (\tau_1 + \tau_2)}$, with $\tau_i$ randomly chosen from
the uniform distribution with support $[0, c_3 \Delta_\beta
]$, for some constant $c_3 >0$. Each random evolution effectively
simulates a measurement of $\ket{\psi_{\beta_r}} $. Due to the
quantum Zeno effect, this random process drives
$\ket{\psi_{\beta_0=0}}$ towards $\ket{\psi_{\beta_m}}$,
with high probability.  The average cost of this method is
\begin{equation}
\langle {\cal T} \rangle = O(n^2/   \Delta   )\label{somma:eq:4}\;,
\end{equation}
where $\Delta$ is a lower bound on $\min_\beta \Delta_\beta$, and we
have used again the assumption $\beta_m=O(n)$, so that the total
number of steps is $m = O(n^2)$. A bound for $\Delta_{\beta}$ gives a
total cost analogous to that of the previous paragraph.

Another QA method to
prepare the desired state,
often used to find the ground states of Ising models,
considers lowering a transverse field from a large initial value. 
In Ref.~\cite{somma:QA3}, the Hamiltonians are of the form
$ \hat E - \gamma \sum_{l=0}^{n-1} \sigma_{\bf x}^l$,
where $\gamma$ is the magnitude of a transverse field and $\hat E$ is the
diagonal matrix that encodes the energies of the states
of a classical Ising model (i.e., the objective function).
This method can be modified to prepare $\ket{\psi_{\beta_m}}$, for $\beta_m < \infty$.
To show this, we invoke
the mapping of Eq.~(\ref{somma:eq:QCM}) and consider  
Eq.~(\ref{somma:eq:Qmetropolis}). Then, the Hamiltonians
\begin{equation}\label{somma:eq:h_gamma}
H_\gamma =  \sum_{l=0}^{n-1} \chi \exp(\beta_m \hat E^l) - \gamma \, \sigma_{\bf x}^l
\end{equation}
can be used to prepare $\ket{\psi_{\beta_m}}$ on a quantum device by adiabatic state 
transformations, using a suitable $\gamma(t)$.
The condition for the transverse field is $\gamma({\cal T}) = \chi$,
so that $H_{\gamma({\cal T})}=H_{\beta_m}$ in that case, as in Eq.~(\ref{somma:eq:Qmetropolis}).
In Refs.~\cite{somma:qmap3,somma:qmap1,somma:qmap2,somma:qmap2p}, we referred to this ``finite-temperature''
extension of the method of Ref.~\cite{somma:QA3} as {\em Extended Quantum Annealing} (EQA).
Note that, since the Hamiltonians in Eq.~(\ref{somma:eq:h_gamma}) do not suffer from the so-called sign problem,
classical quantum Monte Carlo implementations of the EQA are also possible 
\cite{somma:qmap3}.

We use again the quantum adiabatic approximation 
to obtain a suitable $\gamma(t)$. To this end, we 
need a lower bound on $\Delta_\gamma$, the spectral gap of $H_\gamma$.
Considering the worst case instances and under the assumption $E=O(n)$, this gap
satisfies $\Delta_{\gamma} = \Omega((\gamma/c_4)^n)$, for some $c_4> \gamma$~\cite{somma:qmap1}. 
The adiabatic approximation implies
\begin{equation}
\gamma(t) = O(t^{-1/(2n-1)})\label{somma:eq:2} \; .
\end{equation}
Our result on the rate of change of $\gamma$ coincides with that of
Refs.~\cite{somma:QAconvergence,somma:mathfoundations}. 
The total evolution time $\cal T$ can be obtained from $\gamma({\cal
  T})=\chi$. Because  $\chi = \exp ( -\beta_m \kappa)/n$, the scaling of
the total
evolution time as a function of $\beta_m$ and $n$ for the choice of
Eq.~(\ref{somma:eq:h_gamma}) is the same as that of SA in
Eq.~(\ref{somma:eq:geman_geman}).

We consider now the more standard QA method for Ising models mentioned above, where the Hamiltonians are
of the form
\begin{equation}
\label{somma:eq:QAIsing}
\widehat{H}_\gamma=    \hat E  -\gamma(t) \sum_{l=0}^{n-1} \sigma_{\bf x}^l \; ,
\end{equation} 
and the goal is to prepare the ground state of $\hat E$, as opposed to the Gibbs distribution 
for the inverse temperature $\beta_m$.  The scaling  result on the rate of change of $\gamma(t)$ found in
Refs.~\cite{somma:QAconvergence,somma:mathfoundations} coincides with that
of Eq.~(\ref{somma:eq:2}). There is no dependence on $\beta$   when implemented in a
quantum device as a closed system quantum evolution. The final value of the transverse field
is $\gamma({\cal T}) = O(1/n)$, which  guarantees that the optimal configuration is found
with high probability after measuring the ground state of $\widehat{H}_{\gamma({\cal T})}$. This follows from
perturbation theory, assuming that the spectral gap of $\hat E$ (i.e., the difference between the 
two smallest and different $E(\sigma_i)$) is $\Omega(1)$, i.e.,  bounded by a constant. 
The corresponding worst-case bound for the total evolution time is ${\cal T}=O(\exp (c_5 n \log n))$ in this case,
for some constant $c_5>0$. This bound   is still worse than that for the complexity of exhaustive search, but
is better than the absolutely worst case bound obtained for SA
resulting from Eq.~(\ref{somma:eq:geman_geman}), and analogous to the
bound for SA corresponding to a low energy spectrum with a
combinatorial number of elementary excitations. 
As remarked above, and similar
to the results for SA, it is expected that in most practical
cases much faster evolutions would suffice to solve the optimization problem.

Clearly, there are many Hamiltonian paths and corresponding methods
that can be used to prepare $\ket{\psi_{\beta_m}}$ by means of QA.  In
the next section, we construct one such Hamiltonian path that yields a
QA method to prepare the desired state with \textit{provably} improved
complexity than that given by the spectral gap bound for SA.

%


\section{Spectral gap amplification}
\label{somma:sec:3}

We  show how to construct a Hamiltonian path with a
spectral gap $\sqrt{\Delta_\beta}$ when given a SA algorithm where the stochastic matrices have
a gap $\Delta_\beta$. 
This will result in a
quadratic quantum speedup for SA in terms of the spectral gap, which can be exponentially small
in the problem size for hard instances of optimization problems.
Our result is a generalization of
Ref.~\cite{somma:qmap3}, and the   construction in this paper significantly simplifies
the one used for Ref.~\cite{somma:qmap3}.
We consider again the classical-to-quantum mapping of
Eq.~(\ref{somma:eq:QCM}). When the DBC is satisfied by $S_\beta$,
the Hamiltonian of Eq.~(\ref{somma:eq:QCM}) can also be decomposed
as a ``frustration-free'' Hamiltonian. This decomposition
will allow us to use the spectral gap
amplification technique of Ref.~\cite{somma:gapamplification}\footnote{
For completeness, a Hamiltonian $H$ is frustration free if it can be written as $H =
\sum_{k} a_k \Pi_k$, with (known) $a_k \ge 0$, and $(\Pi_k)^2 =
\Pi_k$ projectors.  Further, if $\ket \psi$ is a ground state of $H$, then $\Pi_k \ket \psi = 0 \ \forall
\ k$. We assume $a_k \le 1$.}.

For a Markov process with DBC, we define a corresponding
undirected graph $G$ as the graph with a vertex for each
configuration and an edge for each pair
\begin{equation}
(\Pr \!_\beta  (\sigma_i|\sigma_j),
\Pr \!_\beta (\sigma_j|\sigma_i))\;.
\end{equation}
For each edge we define an unnormalized state
\begin{equation}
  \ket{\mu_{\beta}^{(\sigma_i,\sigma_j)}} = \sqrt{\Pr\!_\beta (\sigma_i|\sigma_j)}
    \ket{\sigma_j} - \sqrt{\Pr \!_\beta (\sigma_j|\sigma_i)}\ket{\sigma_i}\;.
\end{equation}
Note that, from DBC, $\inner{\mu_{\beta}^{(\sigma_i,\sigma_j)}}{\psi_\beta} =
0$. We also define the operators
\begin{equation}
  O_{\beta}^{(\sigma_i,\sigma_j)} = \ket{\mu_{\beta}^{(\sigma_i,\sigma_j)}} 
  \bra{\mu_{\beta}^{(\sigma_i,\sigma_j)}} \ge 0\;.
\end{equation}
Then, Eq.~(\ref{somma:eq:QCM}) can be written as
\begin{equation}
\label{somma:eq:FF1}
H_\beta =  \sum_{(\sigma_i, \sigma_j)} O_{\beta}^{(\sigma_i,\sigma_j)} \; .
\end{equation}
This is a frustration free representation of $H_\beta$, as it is given
by a sum of projectors, and each projector acts trivially on the ground state.

To apply the gap amplification technique of
Ref. ~\cite{somma:gapamplification} efficiently, we need to reduce the number of
operators in the frustration-free representation of $H_\beta$. [As is, the sum over $\sigma_i,\sigma_j$
in Eq.~(\ref{somma:eq:FF1}) involves an exponentially large number of terms.]
We can then assume a given edge coloring of the graph $G$
with   edge chromatic number $q$. It suffices to assume that
the graph $G$ has   degree $D$, which gives a chromatic number at
most $D+1 \ge q$~\cite{somma:vizing1964}. The degree $D$ is determined
by the stochastic matrix, as the edges of $G$ are determined by the nonzero transition probabilities.
For example, for the Glauber dynamics discussed in Sec.~\ref{somma:sec2.1},
$\Pr_\beta(\sigma_i | \sigma_j) \ne 0$ if both configurations differ by a single spin flip. Then,
$D=n$ in this case.
Let
$z_1,\ldots,z_q$ be the different colors. All the operators
$ O_{\beta}^{(\sigma_i,\sigma_j)}$ belonging to one of the colors are, by construction,
orthogonal to each other as they don't share a vertex. That is,
$\tr[O_{\beta}^{(\sigma_i,\sigma_j)}O_{\beta}^{(\sigma_{i'},\sigma_{j'})}]=0$ for $i \ne i'$, $j \ne j'$.
For each $k \in
\{1, \ldots, q \}$, we define the Hermitian operators
\begin{equation}
O_{\beta,k} = \sum_{
    (\sigma_i,\sigma_j) \in z_k 
}  O_{\beta}^{(\sigma_i,\sigma_j)} \; .
\end{equation}
Then,
$H_\beta = \sum_{k=1}^{q} O_{\beta,k} $ is a frustration-free representation. Note that there are many other ways to obtain
a frustration-free representation, as the operators $ O_{\beta}^{(\sigma_i,\sigma_j)}$
can be combined in several ways (i.e., other definitions for $O_{\beta,k} $ are possible).

Given a Hamiltonian $H$ with ground state $\ket \psi$ and gap
$\Delta$, the goal of gap amplification is to find a new Hamiltonian
$\tilde H$ with eigenstate $\ket \psi \ket \nu$ (not necessarily the
ground state) and gap $\tilde \Delta > \Delta^{(1-\varepsilon)}$ with
$\varepsilon>0$. The state $\ket \nu$ is a fixed simple state. In
addition, the implementation complexity of $e^{i \tilde H t}$ must be
of the same order as the implementation complexity of $e^{i H t}$. The
implementation complexity can be defined rigorously as the scaling of
the number of quantum gates necessary to simulate this evolution in a
universal quantum computer. This avoids trivial cases, like scaling
the energies in $H$ by a constant.

We now outline the gap amplification of $H_\beta$ using a
technique that applies to general frustration free
Hamiltonians. We define the Hamiltonian
\begin{equation}
\label{somma:eq:Xdef}
 A_\beta = \sum_{k=1}^q \sqrt{O_{\beta,k}} \otimes (\ket k \bra 0 + \ket 0 \bra k)\;.
\end{equation}
The first property to notice is that  $\bra{\psi_\beta0} A_\beta^\dagger 
A_\beta \ket{\psi_\beta{0}} =0$ and $\ket{\psi_\beta{0}}$ is an eigenstate
in the null space of $A_\beta$.

We label the eigenvalues of $H_\beta$ by
$\lambda_j$, $j=0,\ldots,N-1$, and $\ket{\lambda_j}$ are the eigenstates. We assume
$\lambda_0=0 < \lambda_1 \le \ldots \le \lambda_{N-1}$, so that $\ket{\psi_\beta} = \ket{\lambda_0}$.
Assume now $\lambda_j \ne 0$, and consider the action of
$A_\beta$ on the state $\ket{\lambda_j 0}$, that is,
\begin{equation}
  A_\beta \ket{\lambda_j 0} =  \sum_k \sqrt{O_{\beta,k}} \ket{\lambda_j k}\;.  
\end{equation}
Notice that
\begin{equation}
  \bra{\lambda_j 0} A_\beta \ket{\lambda_j 0} =   \sum_k \bra{\lambda_j k}  \sqrt{O_{\beta,k}} \ket{\lambda_j 0} = 0 \; ,
\end{equation}
and that
\begin{equation}
  \|A_\beta \ket{\lambda_j 0} \|^2 = \sum_k \bra{\lambda_j k} O_{\beta,k}
    \ket{\lambda_j k} =\lambda_j \;.  
\end{equation}
We denote by
\begin{equation}
  \ket{\perp_j} = \frac 1 {\sqrt{\lambda_j}} A_\beta \ket{\lambda_j 0}   
\end{equation}
the corresponding normalized state. Next, note that
\begin{equation}
  A_\beta \ket{\perp_j} = \frac 1 {\sqrt{\lambda_j}} \sum_k O_{\beta,k}
  \ket{\lambda_j 0} = \sqrt{\lambda_j} \ket{\lambda_j 0}\;.
\end{equation}
Therefore, the Hamiltonian $A_\beta$ is invariant in the subspace
$\mathcal{V}_j = \{\ket{\lambda_j 0} ,\ket{\perp_j}\}$. Define $(A_\beta)_j =
(A_\beta)_{|\mathcal V_j}$ the projection of $A_\beta$ in this invariant
subspace. In the basis $\{\ket{\lambda_j 0}, \ket{\perp_j}\}$ we can write
\begin{equation}
  (A_\beta)_j = \left(\begin{array}{cc}
      0 & \sqrt{\lambda_j} \\
      \sqrt{\lambda_j} & 0
    \end{array}\right)\;,
\end{equation}
with eigenvalues $\pm \sqrt{\lambda_j}$. 

We note that, because the $\ket{\lambda_j}$ form a complete basis, $A_\beta$ acts 
trivially on any other state orthogonal to $\oplus_j {\cal V}_j$, and therefore the 
eigenspace of eigenvalue $0$ of $A_\beta$ is degenerate. Although this step is unnecessary, 
we can avoid this degeneracy by
  introducing a ``penalty term'' to change the energies of the undesired eigenstates
in such eigenspace. We then define
 the  Hamiltonian
\begin{equation}
\label{somma:eq:Hfinal}
  \tilde H_\beta = A_\beta + \sqrt {\Delta_\beta} (\openone - \ket 0 \bra 0)\;.
\end{equation}
The subspace orthogonal to $\oplus_j {\cal V}_j$ acquires eigenvalue $\sqrt {\Delta_\beta}$. 
Each space ${\cal V}_j$ is still an invariant subspace of $\tilde H_\beta$, and
\begin{equation}
  (\tilde H_\beta)_j = \left(\begin{array}{cc}
      0 & \sqrt{\lambda_j} \\
      \sqrt{\lambda_j} & \sqrt {\Delta_\beta}
    \end{array}\right)\;.
\end{equation}
The minimum eigenvalue of each of these operators is   $\Omega(\sqrt{ \Delta_\beta})$, 
which is also a bound on the relevant spectral gap of $\tilde H_\beta$\footnote{Note that
the definition of $\tilde H_\beta$ depends on $\Delta_\beta$, which is usually unknown. However,
$\Delta_\beta$ can be replaced by its lower bound $\Delta$ in Eq.~(\ref{somma:eq:Hfinal}), still assuring a
spectral gap of order $\sqrt{\Delta}$.}.

For example, for the Hamiltonian $H_\beta$ of Eq.~(\ref{somma:eq:Qmetropolis}), which is already
expressed as a frustration-free Hamiltonian, we can write
\begin{equation}
\label{somma:eq:Qmetropolis2}
\tilde H_\beta = \sum_{l=0}^{n-1} \sqrt{\chi \exp(\beta \hat E_l) - \chi \sigma_{\bf x}^l } \otimes (\ket{l+1}\bra 0 + \ket 0 \bra{l+1}) +
  \sqrt {\Delta_\beta} (\openone - \ket 0 \bra 0)\;.
\end{equation}

In summary, the state $\ket{\psi_\beta  0}$ is  generally the unique
eigenstate of eigenvalue $0$ of $\tilde H_\beta$. This eigenvalue is
separated by a gap of order $\sqrt{\Delta_\beta}$.
Also, the implementation complexity of evolutions with $\tilde H_\beta$ is similar
to that of evolutions with $H_\beta$, under some reasonable assumptions. For the example of Eqs.~(\ref{somma:eq:Qmetropolis}) and
~(\ref{somma:eq:Qmetropolis2}), if $\hat E_l$ is a ``local'' operator that acts on a few spins, 
evolutions under $H_\beta$ or $\tilde H_\beta$ can be efficiently implemented using known
Trotter approximations or more efficient methods~\cite{somma:efficientsimulation}.

The implication of the spectral gap amplification technique is that
the adiabatic approximation applied to prepare the (excited) state
$\ket{\psi_\beta 0}$ results in a better scaling with
$\Delta_\beta$ than in the case of $H_\beta$.  To show this, we can
use evolution randomization~\cite{somma:AST2,somma:phaserandom2},
which as explained in Sec.~\ref{somma:sec2.2} is a rigorous version of
the adiabatic approximation. The only change with respect to
Sec.~\ref{somma:sec2.2} is that the gap has been improved to $\sqrt
{\Delta_\beta}$.  This results in a quadratic speedup over SA with
respect to the gap $\Delta_\beta$, which is normally the dominating
factor in the analytical bound for the cost of SA, as seen in
Sec. ~\ref{somma:sec2.2}.  The dependence of $\langle {\cal T} \rangle
$ on the error probability can be made fully logarithmic by repeated
executions of the algorithm.  We refer to
Refs.~\cite{somma:gapamplification,somma:AST2,somma:phaserandom2} for
more details about spectral gap amplification and evolution
randomization.

To show that spectral gap amplification
results in a provable quantum speedup, 
we can consider Grover's search problem.
In this case, $E(\sigma_i)=1$ for a particular, unknown
$\sigma_i$, and $E(\sigma_j)=0$ for all $\sigma_j \ne \sigma_i$.
Classical algorithms to find $\sigma_i$ have complexity $\Omega(N)$, i.e.,
they require evaluating the objective function in about half of the inputs, on average.
It is possible to design a SA algorithm, whose stochastic matrices
have gaps $\Delta_\beta=\Omega(1/N)$ to solve this problem.
Then, spectral gap amplification results in a QA method
that outputs $\sigma_i$, with high probability, and has complexity $O(\sqrt N)$,
improving quadratically upon the best classical algorithms.

%
%

\section{Exponential speedup by quantum annealing}
\label{somma:sec:4}
\begin{figure}[htb]
\centering
\resizebox{0.65\textwidth}{!}{
 \includegraphics{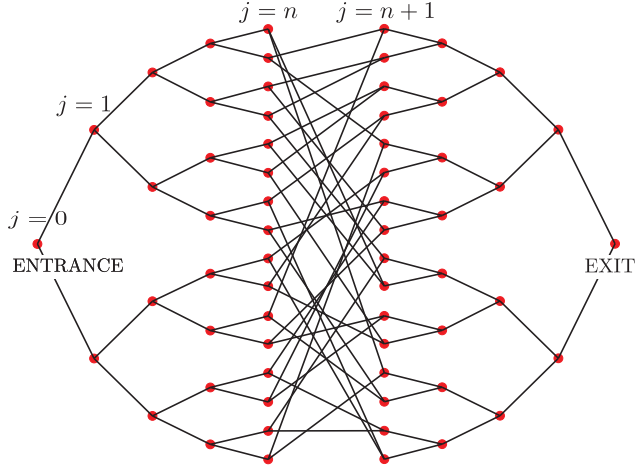}
}
\caption{(From~\cite{somma:expspeedup}.) Two binary trees of depth $n=4$ glued
randomly. The number of vertices is $N = 2^{n+2}-2$. Each vertex
is labeled with a randomly chosen $2n$-bit string. $j$ is the column
number.}
\label{somma:fig:gluedtrees}       
\end{figure}

In this section, we first summarize our results in
Ref.~\cite{somma:expspeedup}, where we considered the problem from
Ref.~\cite{somma:gluedtrees} and   solved it using QA.  We are
given an oracle that consists of the adjacency matrix $A$ of two
binary trees that are randomly ``glued'' (by a random cycle) as in
Fig.~\ref{somma:fig:gluedtrees}.  There are $N =O( 2^n)$ vertices,
which are named with randomly chosen $2n$-bit strings. The oracle
outputs the names of the adjacent vertices on any given input vertex
name. There are two special vertices, ENTRANCE and EXIT, the roots of
the binary trees. They can be easily identified using $A$ because they
are the only vertices of degree two in the graph. The glued-trees
problem is: Given an oracle $A$ for the graph and the name of the
ENTRANCE, find the name of the EXIT. An efficient method based on
quantum walks can solve this problem with constant probability, while
no classical algorithm that uses less than a subexponential (in $n$)
number of oracles exists~\cite{somma:gluedtrees}. Still, a direct QA
method for this problem was unknown. We will then present a QA
approach that efficiently outputs the name of the EXIT with
arbitrarily high probability (in the asymptotic limit).

We let $a(V) \in \{0,1\}^{2n}$ be the bit string that labels vertex $V$.
We assume a Hamiltonian version of the oracle so
that evolutions under $A$ can be implemented.
Our Hamiltonian path for QA is given by ($0 \le s \le 1$)
\begin{equation}
\label{somma:eq:QApath}
H(s) = (1-s) H_{\rm ENTRANCE} - s(1-s) A + s H_{\rm EXIT} \; ,
\end{equation}
with 
\begin{eqnarray}
H_{\rm ENTRANCE} | a({\rm ENTRANCE}) \rangle &=&- \alpha | a({\rm ENTRANCE}) \rangle \; , \nonumber\\ 
  H_{\rm EXIT}  | a({\rm EXIT}) \rangle  &=&- \alpha | a({\rm EXIT}) \rangle \; ,
\end{eqnarray}
and any other eigenvalues of these Hamiltonians are zero. $\alpha >0$ is a constant, e.g., $\alpha=1/\sqrt 8$
works.

\begin{figure}[htb]
\centering
\resizebox{0.75\textwidth}{!}{
 \includegraphics{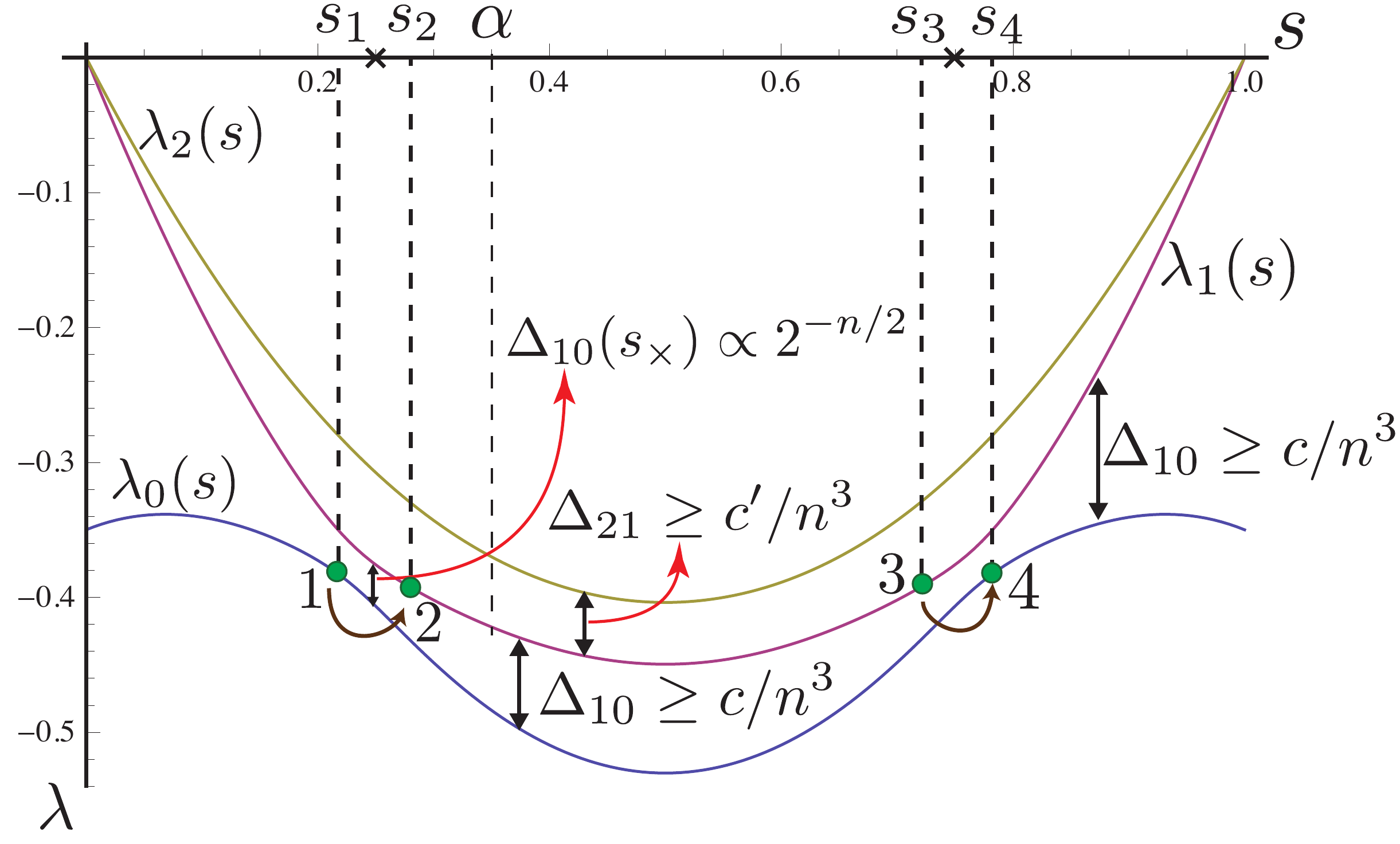} 
}
\caption{(From~\cite{somma:expspeedup}.) Eigenvalues and energy gaps of $H(s)$, and scaling with 
the problem size $n$. $s_\times=\alpha/\sqrt 2$ and $\alpha=1/\sqrt 8$ in this case.}
\label{somma:fig:gaps}       
\end{figure}

From Eq.~(\ref{somma:eq:QApath}), it is clear that when $s=1$, the ground state of $H(1)$ is
$| a({\rm EXIT}) \rangle$. A measurement of this state allows us  to obtain the solution
to the glued-trees problem. The QA method is then designed to evolve the ground
state of $H(0)$ towards that of $H(1)$. One may attempt to do this adiabatically. Nevertheless,
the spectrum of $H(s)$, depicted in Fig.~\ref{somma:fig:gaps}, shows that the 
smallest spectral gaps between the two lowest-energy
states can be exponentially small in the problem size. This seems to imply that 
the total time required to adiabatically prepare $| a({\rm EXIT}) \rangle$ 
from $| a({\rm ENTRANCE}) \rangle$ would be exponentially large,
resulting in an inefficient state preparation.

Remarkably, if $s$ is changed in time so that it satisfies $\dot s (t) \propto 1/n^6$,
the state $| a({\rm EXIT}) \rangle$ can be prepared from $| a({\rm ENTRANCE}) \rangle$
with arbitrary low error (in the asymptotic limit), thus solving the
glued-trees problem efficiently. Such an annealing schedule for $s$
implies that the initial ground state is eventually transformed into the first excited state
for $s \ge s_\times$.
But, at a later time, the symmetries of the spectrum around $s=1/2$ imply that the 
first excited state is transformed back to the lowest-energy state (for $s \ge 1-s_\times$), eventually preparing
$| a({\rm EXIT}) \rangle$ when $s \rightarrow 1$. The evolution of the state is 
sketched in Fig.~\ref{somma:fig:gaps}, where points (1,2) and (3,4) depict where the 
diabatic transitions between the two lowest-energy states occur. 
The details of the calculations for the spectral properties of $H(s)$, as well as the
 details about the efficient QA solution are given in Ref.~\cite{somma:expspeedup}.

It is important to note that, in the glued-trees problem, all the {\em interesting}
quantum dynamics occurred in the manifold given by the two lowest-energy states.
This additionally implies that, if when $s \rightarrow 0$ either state is prepared
with probability $1/2$,   the state $| a({\rm EXIT}) \rangle$ would also be prepared
with probability of almost $1/2$ by following an annealing schedule in which $\dot s (t) \propto 1/n^6$.
The simple reasoning is that such a manifold is adiabatically decoupled from other excited
states for that choice of $s(t)$. That is, the spectral gap between the first and second excited
states, $\Delta_{21}$, is of order $1/n^3$ rather than exponentially small for $1>s>0$-- see Fig.~\ref{somma:fig:gaps}.
 Thus, randomization of initial state preparation would
also provide an efficient QA method to solve the glued trees problem, with probability of almost $1/2$.

One may wonder how general
this efficient approach is for solving other optimization problems. Motivated by Ref.~\cite{somma:expspeedup}, 
in Ref.~\cite{somma:max2sat} the authors considered recently different QA
evolutions to solve the well-known MAX 2-SAT problem. In this case,
one is given a Boolean formula in conjunctive normal form, such that each clause
contains at most two variables. The goal is two find an assignment to the variables
such that a maximum number of clauses is satisfied. MAX 2-SAT is a NP-hard problem
as one can reduce the well known NP-complete problem 3-SAT into MAX 2-SAT.
The Hamiltonians used in the evolution act on a system of $n$ qubits and are parametrized as
$H(s) = (1-s) H_B+ s H_P$ with
\begin{eqnarray}
\hspace*{-0.5cm}
H_B &=& -\sum_{l=1}^n (1-\sigma_{\bf x}^l)/2 \; , \\
H_P&=& - \hspace*{-0.3cm}\sum_{z \in \{0,1\}^n} f(z) | z \rangle \langle   z| \; ,
\end{eqnarray}
where $\sigma_{\bf x}^i$ is the Pauli spin flip operator acting on the $l$th qubit. $f(z)$
is the sum of the clauses in the Boolean formula on input $z$. Thus, a measurement on the ground state of
$H_P$ gives the solution to the corresponding MAX 2-SAT instance.

The ground state of $H_P$ can be prepared using QA, evolving $s$ from $0$ to $1$.
The authors of Ref.~\cite{somma:max2sat} note that, for hard instances 
(according to a particular method that separates hard from easy instances)
preparation of the ground state of $H_P$ would take an extremely long time (simulations where ran for $n=20$).
Nevertheless, for those same instances, the probability of success in preparing the ground state is enhanced
as one increases the rate of change of $s$. The nature of such an enhancement is similar to that of the glue-trees
problem, where the QA evolution also generates diabatic transitions between different eigenstates.
This opens a new door for the development of fast quantum algorithms, based on the idea of QA, 
for discrete optimization.

\section{Conclusions}
\label{somma:sec:conclusions}
We presented three strategies to solve combinatorial optimization
problems based on the idea of quantum annealing. All our strategies were
developed in the context of quantum computation or quantum simulation, and will require
a coherent quantum device for their implementation. Some strategies provide (provable) polynomial and exponential
quantum speedups with respect to the corresponding classical methods for certain problems. 

To obtain the results, we devised particular Hamiltonian paths and
techniques so that the corresponding quantum evolution prepares a
desired quantum state faster than the corresponding classical
algorithm. First, we mapped the stochastic matrix of a classical
Markov process into a quantum (frustration-free) Hamiltonian, whose
ground state encodes the Gibbs (equilibrium) distribution of the
classical system.  In one case, we used the idea of spectral gap
amplification, which is basically a mapping that takes a
frustration-free Hamiltonian and efficiently outputs another
Hamiltonian, having a much larger spectral gap, but preserving the
ground state as eigenstate.  A larger gap implied a (provable) faster
way to prepare the desired quantum state adiabatically. In another
case, we used the idea of diabatic transitions, allowing us to prepare
a ground state by traversing other higher-energy states.  Our
techniques may be generalized to other problems; a step in this
direction was recently given in Ref.~\cite{somma:max2sat}.

%
%
%

\end{document}